\documentclass{osa-article}
\journal{ome}
\articletype{Research Article}

\begin{document}

\title{Modular system for fluorescence-based single photon generation using retro-reflector}

\author{Hee-Jin Lim,\authormark{1} Kwang-Yong Jeong,\authormark{2} Dong-Hoon Lee,\authormark{1} and Kee Suk Hong\authormark{1,*}}

\address{\authormark{1}Korea Reasearch Institute of Standards and Science (KRISS), Daejeon 34113, South Korea\\
\authormark{2}Department of Physics, Korea University, Seoul 02841, South Korea}

\email{\authormark{*}hongi2011@kriss.re.kr}

\begin{abstract}
Apparatus for fluorescence-based single photon generation includes collection optics and various setups for characterization.
Managing this system often reveals complexity in such a way that adjusting in a small region changes optimal alignments of others.
We suggest here a modular system, where the optimal alignment is given to each compartment and tested independently.
Based on this concept, we built a system for single photon generation with fluorescence center in hexagonal boron nitride nano-flake, advantageous for scaling up the number of single mode fiber output and a high degree of stability.
The system allowed for a practical use of single photon stream extended over an hour with a uniform count rate of small fluctuation levels.
\end{abstract}

\section{Introduction}
Single photon generation with fluorescence centers has been intensively performed with nano-materials like crystal defect in hexagonal boron nitride nano-flake and silicon vacancy color center in nano-diamond,\cite{Neu:2011ik,choi:2014apl,Tran:2016jw} and exhibited affordable count rates at room temperature.
Related setups commonly include a microscope for optical collections, and various modules to characterize emission properties like photoluminescence spectrum and photon number statistics. 
For practical use of single photons, the whole system requires a high degree of stability both for maintaining a rate of photon generation, and for preserving alignments.
However, complexity arises from the interdependence of an optimal alignment in each fraction, such that a small change in a part of optics alters the optimal condition of other.
This can cause a trouble for searching the fastest route to recover from unintended changes of optical alignment.
An idea we tried here is to compartment the optics setup into modules, while condition of optimization is given to each module without conflicts and tested independently.
The self-characterization to the  independently appointed optimal condition is the key idea of modularization.
This is similar to a measurement system made up by various kinds of equipment with a built-in self-test, and protocols needed to regularize signals between them.

\begin{figure}[h!]
	\centerline{\includegraphics[width=8cm]{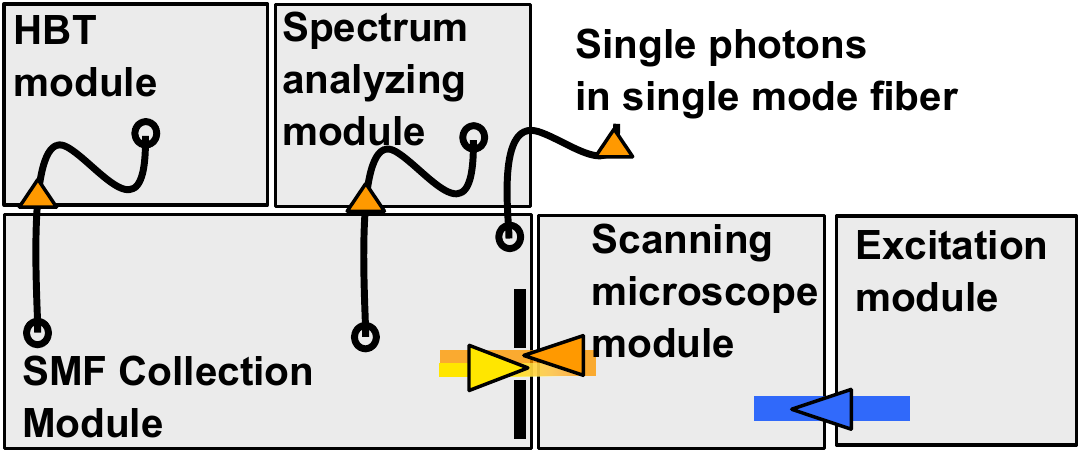}}
	\caption{Outline of a system made up by optical and test modules for single photon generation. One of single mode fiber outputs is reserved for practical applications of single photons. Definition used is HBT: Hanbury Brown-Twiss interferometer for measurements of photon number statistics.
	}
	\label{fig:system}
\end{figure}

The system built under this idea can be outlined as the schematic image shown as Fig. \ref{fig:system},
which is designed to funnel single photon emission from a fluorescence centre through a microscope into a single mode fiber.
The \textit{scanning microscope module} scans a specimen where nano-flakes are dispersed, and fixes at a detection point.
The \textit{single mode fiber (SMF) collection module} plays numerous roles of delivering single photons into a number of SMFs in a switching manner.
The module has three SMF outputs of single photon stream and hand-free optical switches operated in programmable ways; two SMFs are connected to Hanbury  Brown-Twiss (HBT) interferometer and \textit{spectrum analyzing module} for characterizations, and the last channel of SMF is reserved for practical applications of single photon source.
This intends to program a procedure of characterizing and releasing single photons, and allows for spot checks of system in the middle of single photon generation.

Aligning components inside the \textit{SMF collection module} sensitively depends on modules connected via free-space optics, and had taken many efforts to be consistent with the whole system optimally.
However, we do this effortlessly with the help of apparatus for self-tests inside the module, as being delineated in Sec. \ref{sec:SMFcol}.
The condition of alignment for SMF coupling is memorized in the setting of a reference beam internally released by a retro-reflector.
Opposed to simplicity of connection via SMF,\cite{PRL.118.233602,brash2019light} interfacing modules via free-space optics needs the regularization provided with the reference beam of a defined mode released externally.
We devised the reference beam to imitate the signal beam indistinguishably and to optimize directly coupling of modules.
The more detail will be explained in Sec. \ref{sec:system}.
For the demonstration of the basic idea, we confined our considerations of modal properties to Gaussian modes in free-space and the mode of SMF.
However, our design has possibilities to improve collection efficiency with ideas exploiting various types of fiber\cite{4758654,6683058, Sakamoto:11} or nano-structures\cite{PhysRevLett.122.113602,Chen:2018vy,C7NR08249E}.
We discussed them in the final section.

We successfully demonstrated single photon generations in this system.
We examined quality and stability of single photon streams, and whether they are acceptable for precision measurements of quantum radiometry\cite{Rodiek:2017en} in which maintaining a photon count rate is highly important.
For fast demonstrations, we made use of single photon emissions from hexagonal boron nitride (hBN) nano-flake at room temperature\cite{Tran:2016gu}.
Results of quality and stability are presented in Sec. \ref{sec:results}.

%%%%%%%%%%%%%%%%%%%%%%%%%%%%%%%%%%%%%%%%%%%%%%%%%%%%%%%%%%%%%%%%%%%%
%%%%%%%%%%%%%%%%%%%%%%%%%%%%%%%%%%%%%%%%%%%%%%%%%%%%%%%%%%%%%%%%%%%%

\section{Constructing a module with a built-in self-test  for single mode fiber coupling}\label{sec:SMFcol}
\begin{figure}[ht]
	\centerline{\includegraphics[width=13cm]{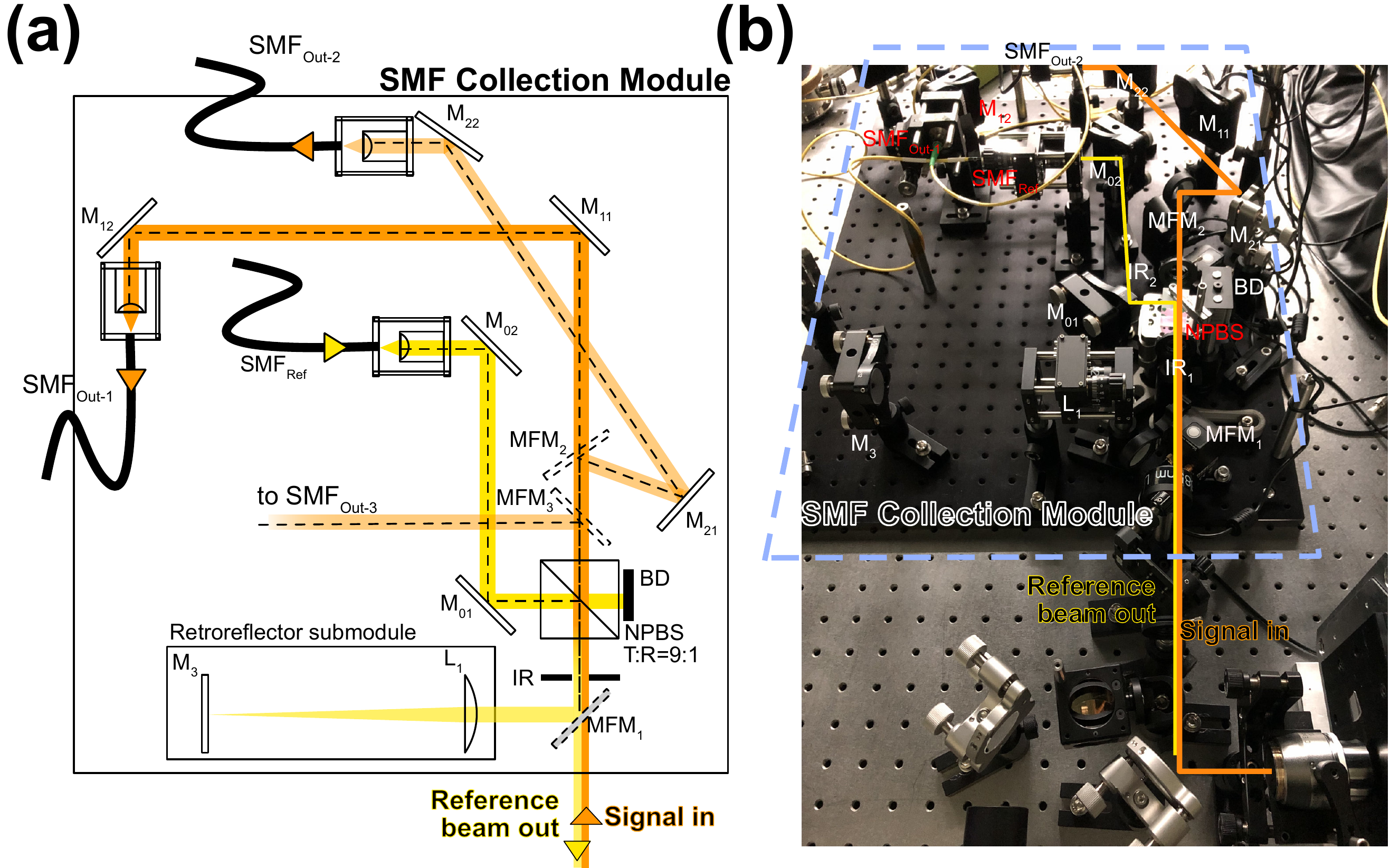}}
	\caption{ 
		\textbf{(a)} Layout of the \textit{single mode fiber collection module} that works for conversions between free-space and fiber optics, and has two fiber outputs (SMF$_{\mathrm{Out-1}}$ and SMF$_{\mathrm{Out-2}}$). Definitions used are $\mathrm{M}_{ij}$: mirror, $\mathrm{MFM}_i$: motorized flip mirror, L$_i$: lens, NPBS: non-polarizing beam splitter, BD: beam dump, IR$_i$: iris. 
		\textbf{(b)} Picture of the module used for a confocal scanning microscope. The third SMF output (SMF$_{\mathrm{Out-3}}$) is out of the photo range in \textbf{(b)}, and shorten in \textbf{(a)}.
	}
	\label{fig:module}
\end{figure}

A major role of the \textit{SMF collection module} is to funnel a propagation beam from a microscope into a SMF.
We consider the lowest-order mode ($\mathrm{LG}_{00}$) for compatibility, because this is transformed well to the fundamental mode ($\mathrm{HE}_{11}$) of SMF  by collimation optics.\cite{Forbes:2015fo}

A novelty of our design presented in Fig. \ref{fig:module} lies in a built-in apparatus for self-test and a reference beam (REF) established to help an external connection.
The apparatus importantly has a retroreflector that reflects the REF back into the module, and emulates a signal beam.
The retroreflector, made of a plano-convex lens ($L_1$) and a mirror ($M_3$), produces a minimum displacement by adjusting $M_3$.\cite{Donley:2005jm}
When a mirror ($\mathrm{MFM}_1$) is placed on by a motorized flip stage, the reference beam imitates a signal beam after being reflected from the retroreflector to the same path.
When this beam is coupled with a SMF optimally, the self-test is complete.
Then the reference beam is released by flipping $\mathrm{MFM}_1$, and able to guide alignments of signals to be collected.

A beam splitter (NPBS) with an asymmetric transmission/reflection (T/R) ratio of 9/1 is meant to transmit a signal beam mostly.
The loss of reflection can be reduced by replacing the NPBS with a pellicle on a flip mount of high repeatability, and removing temporarily from the beam path for low loss operation.
Optical paths are switched by a number of MFMs controlled by computer program, and the module remains hand-off. 
The diameter of the reference beam and the signal beam in free-space is 2.7 mm, determined both by the focal length ($f_{\mathrm{asph}}$, 14 mm) of a collimating aspheric lens (Thorlabs, C560TME-B) and  a mode field diameter  (MFD, $4 \mu$m) of fiber (SM600). 
Such extent is large enough to fill an aperture of objective lens.
Coupling efficiencies into three SMFs (SMF$_{\mathrm{Out-1,2,3}}$ in Fig. \ref{fig:module}), which depends on displacement, angle, and collimation,\cite{Toyoshima:2006cl} were measured to be > 80 $\%$ in most cases.
We used a pair of mirrors to couple into a SMF; one ($M_{12}$) is to control an angle and the other ($M_{11}$) to compensate a displacement.
Mirror mounts (Newport 9812) have a good angular sensitivity around 20 $\mu$rad that is smaller than the maximum deviation allowed for coupling ($\Delta \theta_\mathrm{dev}$, 140 $\mu$rad) given by the following equation;
\begin{equation}
f_{\mathrm{asph}}\times\Delta \theta_\mathrm{dev} < \frac{\mathrm{MFD}}{2}.
\end{equation}
A motorized flip mirror (MFM$_2$, Thorlabs MFF101) is placed to make the signal path connected to the second SMF, and has an acceptable angular repeatability of 50  $\mu$rad $< \Delta \theta_\mathrm{dev}$.

Our first version of the module was built on a plate of the small area 450 $\times$ 300 mm$^2$, and replaced with the present one to increase the number of SMF output.
We have re-used the old one for other systems without re-alignments.
Small adjustment of the angle of M$_{12}$ can be required when the module is moved to a new platform, due to mechanical elasticity of the 10 mm thick plate.
This can be improved by transforming the design to a compact body of high mechanical stability.

To attain a greater robustness of fiber optic coupling, we can suggest single mode photonic crystal fiber (SM-PCF).
Its large MFD can allow a lateral shift of fiber position within few micrometers,\cite{4758654} while high f-number of collimation optics helps to reduce aberrations.\cite{1992aooe11S}
The single mode property suits for the modular design in which modules have to share the same spatial profile of propagation mode.

An imaging camera, made up by a camera sensor (Edmund optics acA2440-20gc, CMOS in Fig. \ref{fig:Confocal}) and an imaging lens  (Thorlabs ACA254-150, $L_2$), was used to compare angles and to find the optimum setting of collimation.
Provided that the pixel size (3.5 $\mu$m) of the camera sensor is smaller than the theoretically attainable beam waist ($w_0$, 5.7 $\mu$m) given by the f-number of $L_2$,
an angle resolution ($\Delta\theta_{\mathrm{res}}$) depends mainly on the diameter of aperture ($D$);\cite[p.676]{siegman1986lasers}
\begin{equation}
\Delta\theta_{\mathrm{res}} =  \frac{2 \lambda}{\pi D}.
\end{equation}
The angle resolution is expected to be in the same order of 50 $\mu$rad for the active $D$ of 10 mm theoretically, and proved to be better than 70 $\mu$rad experimentally. 
This degree of resolution is adequate to inspect the angular alignment required for SMF coupling.

Here is a summary of the procedure to do self-tests.
\begin{enumerate}
\item[a.] Flip MFM$_1$ to bend the reference beam into the retroreflector.
\item[b.] These steps are only for the initial alignment of fiber coupling, and skipped in self-tests.
\begin{itemize}
\item Insert a laser power reversely from the SMF output using any arbitrary fiber laser. 
\item Adjust both M$_{11}$ and  M$_{12}$ to make the collimated beam of the fiber laser parallel to the coming reference beam.
\item Remove the fiber laser.
\item Optimize angles and positions of the reference beam by adjusting M$_{12}$ and M$_{11}$ respectively until the maximum power is attained at the SMF output.
\end{itemize}
 \item[c.] Adjust M$_{12}$ slightly until the collected power of the reference beam is maximized in SMF output.
\item[d.] Flip MFM$_1$ to release the reference beam.
\end{enumerate}

%%%%%%%%%%%%%%%%%%%%%%%%%%%%%%%%%%%%%%%%%%%%%%%%%%%%%%%%%%%%%%%%%%%%
%%%%%%%%%%%%%%%%%%%%%%%%%%%%%%%%%%%%%%%%%%%%%%%%%%%%%%%%%%%%%%%%%%%%

\section{Integration of modules into a system for single photon generation}\label{sec:system}
\begin{figure}[ht]
	\centerline{\includegraphics[width=10 cm]{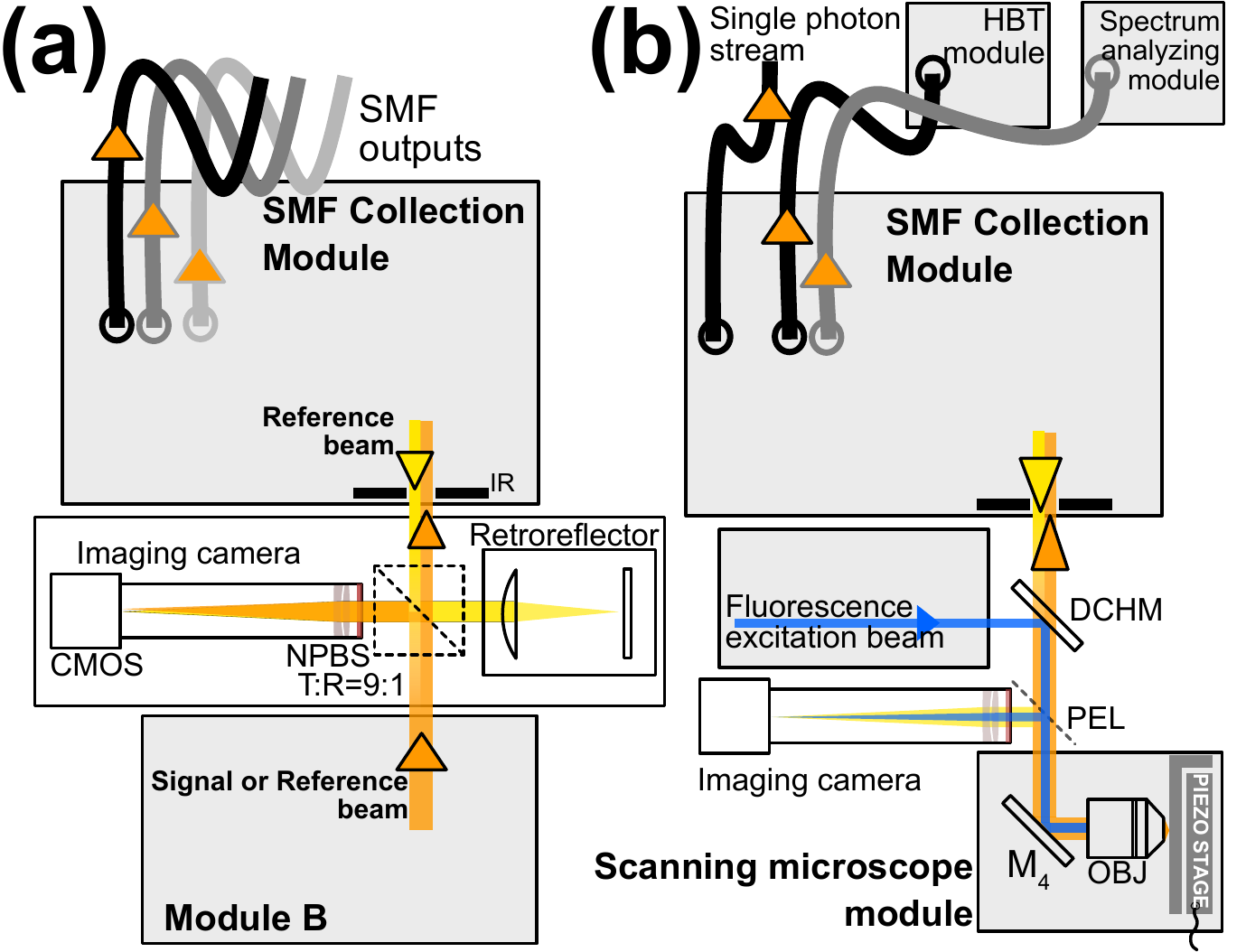}}
	\caption{ 
		\textbf{(a)} General method to connect modules. A setup made up of an imaging camera and retroreflector intermediates connections, comparing angles between two reference beams.
		\textbf{(b)} Modularized confocal scanning microscopy system.
		Definition used are SMF: single mode fiber, DCHM: dichroic mirror, PEL: pellicle, M$_i$: mirror, IR$_i$: iris, OBJ: objective lens, and HBT: Hanbury Brown-Twiss interferometer. }
	\label{fig:Confocal}
\end{figure}

The \textit{scanning microscope module} is integrated with the \textit{SMF collection module} into a confocal scanning microscope (CSM).
Without losing generality, we show a method of connection  between modules in free-space optics, and then particularizes for the current case of CSM.
A free-space connection between arbitrary modules ($A$ and $B$) technically requires to match path and beam waist.
We facilitate a small setup for doing this with reference beams produced by module $A$ and $B$, as shown as Fig. \ref{fig:Confocal}(a).
A beam splitter (NPBS) of 9/1 (T/R ratio) is used to  extract the reference beam from module $B$ and send to an imaging camera.
A retroreflector allows the reference beam of module $A$ to be collected into the imaging camera together with the one from module $B$.
This small setup for mediation can contain two mirrors to match the optics height.
When the condition of alignment is optimized, two spots, which are corresponding to angles of reference beams, are appearing at the same position in the camera.
A beam shift is additionally checked against in an entrance iris (IR$_1$).

The same method was applied to CSM shown as Fig. \ref{fig:Confocal}(b).
An objective lens (OBJ) and a flat sample on the image plane reflects the reference beam into the imaging camera, as if the retroreflector plays in Fig. \ref{fig:Confocal}(a).
Then the process is similar to the one previously explained, except that a dichroic mirror (DCHM) allows for insertion of an excitation beam along the reference beam.
Visual assists of the imaging camera help to coincide a focused spot of the excitation beam with the collection point, reducing angular deviations before the OBJ.
In order that the signal beam has zero displacement with the reference beam, a mirror M$_4$, which is put close to the OBJ and tunes an incidence angle to its pupil, has to be adjusted until the reflected beam has a zero displacement to IR$_1$.
Following two additional steps are unnecessary, but effective to gain a higher level of precision.
(1) An additional mirror (M$_5$) other than M$_4$ can preemptively be placed to correct remained angular mismatches, though we have managed without it in the setup shown in Fig. \ref{fig:module}(b). 
(2) These mirror adjustments (M$_4$ and M$_5$) can be assisted by monitoring the intensity collected in a SMF of module $A$.
To avoid errors of chromatic aberration, the wavelength of reference beam was set at the middle of fluorescence spectrum as a target wavelength.
Then, the focused spot of reference beam is identical to the collection point defined by a conjugate image projected from the SMF core.
This scheme has a diffraction-limited spatial filter 4 $\mu$m in diameter corresponding to MFD of fiber, and configures a confocal microscope.

To estimate a coupling efficiency from sample to SMF, we placed a mirror in the position of sample and collected the reference beam after it passed twice through the OBJ.
The coupling ratio from free-space to SMF was decreased about 63 \% compared to the one measured during self-tests inside the \textit{SMF collection module}.
Given a transmittance of the OBJ as 90 \%, we estimated the collection efficiency to be about 21 \%.
However, the real coupling efficiency assumed for the emission from an electrical point-dipole source requires precise calculations,\cite{Zschiedrich:2018fc} and their results will be smaller than the estimation performed here with a well-defined Gaussian beam.

We employs a 3-axis piezo-stage (Physik Instrumente P-517) to scan positions of fluorescence centers, and acquires photon statistics and spectra.
We acquires maps of fluorescence signals using a commercial data acquisition board (DAQ, NI USB-6361).
The DAQ has a common clock given by a user-determined sampling rate, and executes two things simultaneously; one is controlling voltages to change positions through a servo-controller, and the other is counting photon arrival events signaled by a single photon avalanche counting modules (SPCM, Excelitas SPCM-AQRH).
Photons collected into SMFs are delivered to external modules of Hanbury Brown-Twiss interferometer (HBT) and spectrometer.
The HBT module has two SPCMs and a time tagger (PicoQuant Hydraharp 400) that collects the time period ($\tau$) between two sampled photon arrivals, which are later analyzed into a histogram $g^{(2)}(\tau)$. 
The spectrometer module consists of a highly sensitive spectrum camera (Horiba Synapse) and a monochromator (Horiba iHR320) with homemade f-number matching optics for a SMF input, and can resolve a full-width-half-maximum linewidth as 0.6 nm in wavelength.

Here is a summary of the procedure to setup a confocal microscopy system.
\begin{itemize}
\item Run the \textit{self-test} of  the \textit{SMF collection module}. Release the reference beam.
\item Mount a flat sample with the OBJ, and let the reference beam reflected from the surface of sample.
\item Locate the sample surface at the image plane. Then this group of the OBJ and the sample works like a retroreflector for the reference beam. This step is helped by the iris installed at the entrance of  \textit{SMF collection module}.
\item Find the reflected beam hitting near the hole of the iris.
\item Adjust M$_4$. This action shifts the reflected reference beam laterally, and does not change angles to the iris. Make the reflected beam pass through the iris.  This step compensates a tilt of the OBJ and a resultant displacement of the signal beam.
\item Monitor the power of the reflected reference beam in a SMF output. Adjust both M$_4$ and the sample position along the beam axis slightly until the collected power in the SMF output hits the highest level.
\item Place the PEL to attain images from the camera. Adjust the DCHM to make a focused spot of the excitation beam  superposed on  the reference beam in the image.
\item Remove the PEL to clean the collection path. Turn off the reference beam.
\end{itemize}

%%%%%%%%%%%%%%%%%%%%%%%%%%%%%%%%%%%%%%%%%%%%%%%%%%%%%%%%%%%%%%%%%%%%
%%%%%%%%%%%%%%%%%%%%%%%%%%%%%%%%%%%%%%%%%%%%%%%%%%%%%%%%%%%%%%%%%%%%
\section{Results of single photon generation: quality and stability}\label{sec:results}

\begin{figure}[ht]
	\centerline{\includegraphics[width=8.5cm]{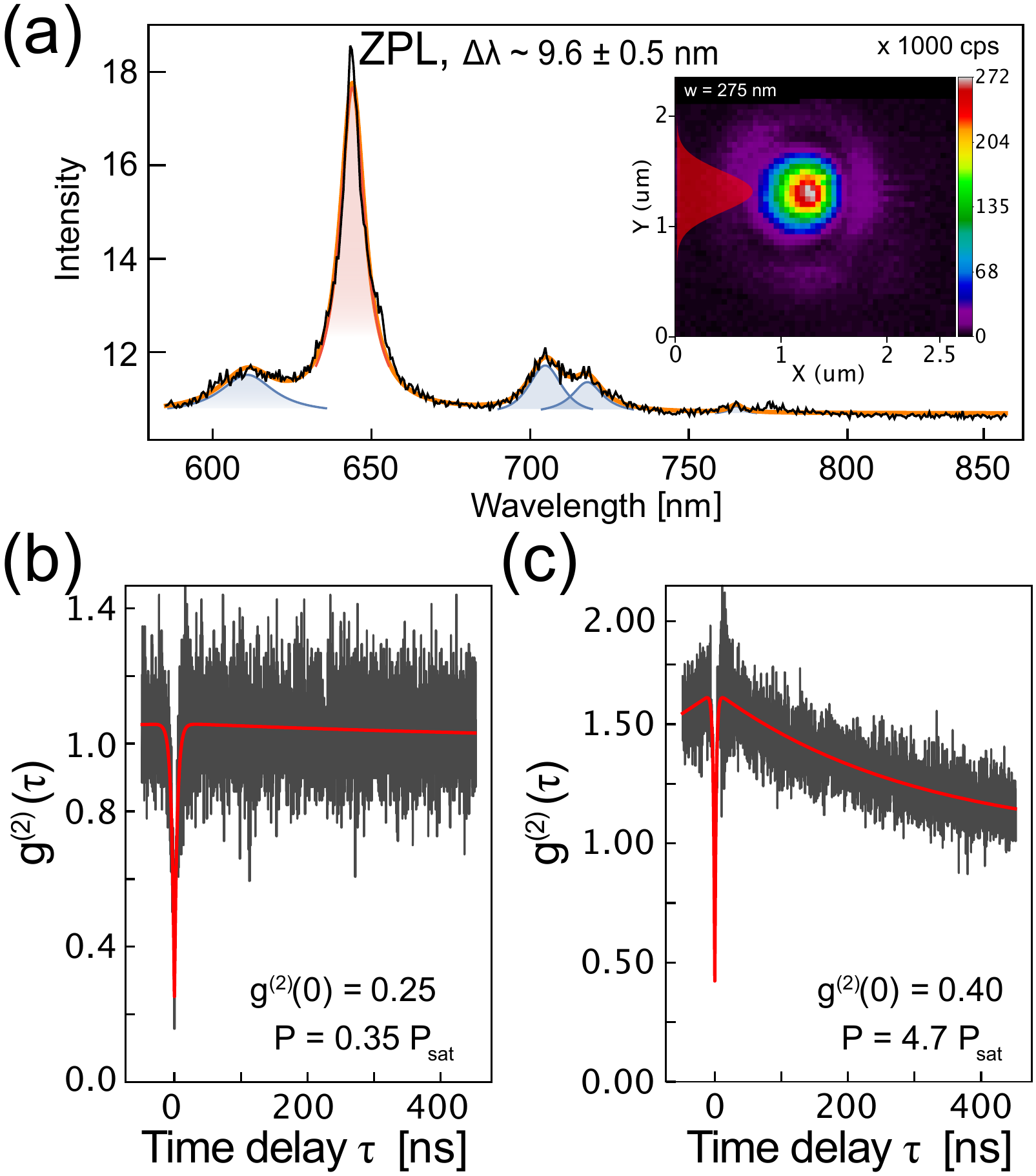}}
	\caption{ 
		\textbf{(a)} Spectrum of photoluminescence collected from a fluorescence center inside hexagonal boron nitride nano-particle.
		An inset shows the intensity map of a fluorescence center measured by position scanning.
		The map is fitted to a model: $I(x) =I_0 \exp (-2 x^2/w^2)$, and a waist ($w$) 275 nm. 
		\textbf{(b,c)} Time correlation histograms of photon detection events collected in the Hanbury Brown-Twiss interferometer module.
		Both histograms were acquired from the same photoluminescence signal, but under different excitation power ($P$) of \textbf{(b)} $0.35\times P_\mathrm{sat}$ 
		%(0.3 mW) 
		and \textbf{(c)} $4.7\times P_\mathrm{sat}$ 
		%(4 mW) 
		compared to the power at a saturation point of count rate ($P_\mathrm{sat}$).
		%, 0.85 mW).
	}
	\label{fig:data}
\end{figure}

To characterize properties of the system, we demonstrated single photon generation, and tested a stability related to maintaining count rates.
We made use of efficient single photon emissions of fluorescence defects in nano-flakes of hexagonal boron nitride (hBN).\cite{Nikolay:2019vl}
Those nano-flakes were spin-coated on an oxidized silicon substrate, and the sample was mounted on the piezo-stage.
We searched for a special defect able to emit single photons among randomly dispersed one by studying photoluminescence at room temperature, which was excited by a diode-pumped-solid-state laser of 532 nm in wavelength.

An inset in Fig. \ref{fig:data}(a) is a map of PL measured with position scanning.
A focused spot of the excitation beam in optical imaging is theoretically expected to have a waist ($w$) of 266 nm due to the diffraction, and expressed by a point spread function with the numerical aperture of the OBJ (NA = 0.85);
\begin{equation}
w = 0.425 \frac{\lambda}{\mathrm{NA}},
\end{equation}
This equation approximates the \textit{Airy} function of paraxial point-spread model\cite[p.93]{novotny2006principles} into the Gaussian model: $I(x) =I_0 \exp (-2 x^2/w^2)$.
Experimentally, the waist of fluorescence center, which we acquired like the inset of Fig. \ref{fig:data}(a), was fitted to be 275 nm, and close to the theoretical limit explained previously, indicating that the nano-particle is well isolated.
The experimental value can be larger due to a difference of wavelengths between the excitation beam and PL signals.

A PL spectrum of this fluorescence defect was resolved into a 9.6 nm-wide zero-phonon-line at 643 nm in wavelength and a few of phonon side-bands, which corresponds to a red-shaded peak and surrounding blue-shaded peaks respectively in Fig. \ref{fig:data}(a).
This feature is the same as those observed in earlier studies,\cite{Jungwirth:2017kb} and the spectral shape indicates that this fluorescence defect has the atomic formation similar to the one found to be radiatively efficient.\cite{Grosso:2017gx}

Correlation dips, appeared at the zero time delay as in Fig. \ref{fig:data}(b) and (c), are the prominent feature of single photons, meaning that single photons are not detected twice.
Photon number statistics ideally have no coincidence of detection events in the HBT module and zero second-order coherence ($g^{(2)}(0)=0$).\cite[p.245]{loudon2000quantum}
Our best result was attained like $g^{(2)}(0)=0.25$ shown as Fig. \ref{fig:data}(b), measured under an excitation power ($0.35 P_\mathrm{sat}$)
%, 0.3 mW
lower than one 
at a saturation point of photon count rate ($P_\mathrm{sat}$).
%, 0.85 mW). 
This value of $g^{(2)}(0)$ implies the 13 \% chance of noise photons\cite{Santori:2004fr,Lee:2016dy} present in PL of the single photon emitter or the background intruded by other fluorescence sources.
With the maximum excitation power of $4.7 P_\mathrm{sat}$,
%(4 mW),
the count rate reached to total $3 \times 10^5$ photons per second maximally and $g^{(2)}(0)$ was increased to $0.4 \pm 0.03$ shown as Fig. \ref{fig:data}(c).
The decreasing slope of $g^{(2)}(\tau)$ for $\tau > 0$ emerged more clearly at the high excitation power.
The origin was explained by relaxations to metastable dark states at which the fluorescence emission is paused until they are released to bright levels.\cite{PhysRevB.69.205324}
The increased $g^{(2)}(0)$ still remains in the regime that the single photon emission still dominates the chance to be trapped into metastable dark state or an emission of noise photons.\cite{Berhane:2017gm}
This result of count rate is smaller than previous reports by an order of difference,\cite{Tran:2016gu,Grosso:2017gx} attributed to difficulties in sample preparations and coupling losses from dipole radiation to the mode in SMF\cite{Zschiedrich:2018fc}.
We had a low chance (one averagely found in a scanning area of $200 \times 200 \mu$m$^2$) to find fluorescence centers emitting single photons, and observed a wide variation of brightness (up to an order of magnitude) from different fluorescence centers in our sample.

\begin{figure}[ht]
	\centerline{\includegraphics[width=8.5cm]{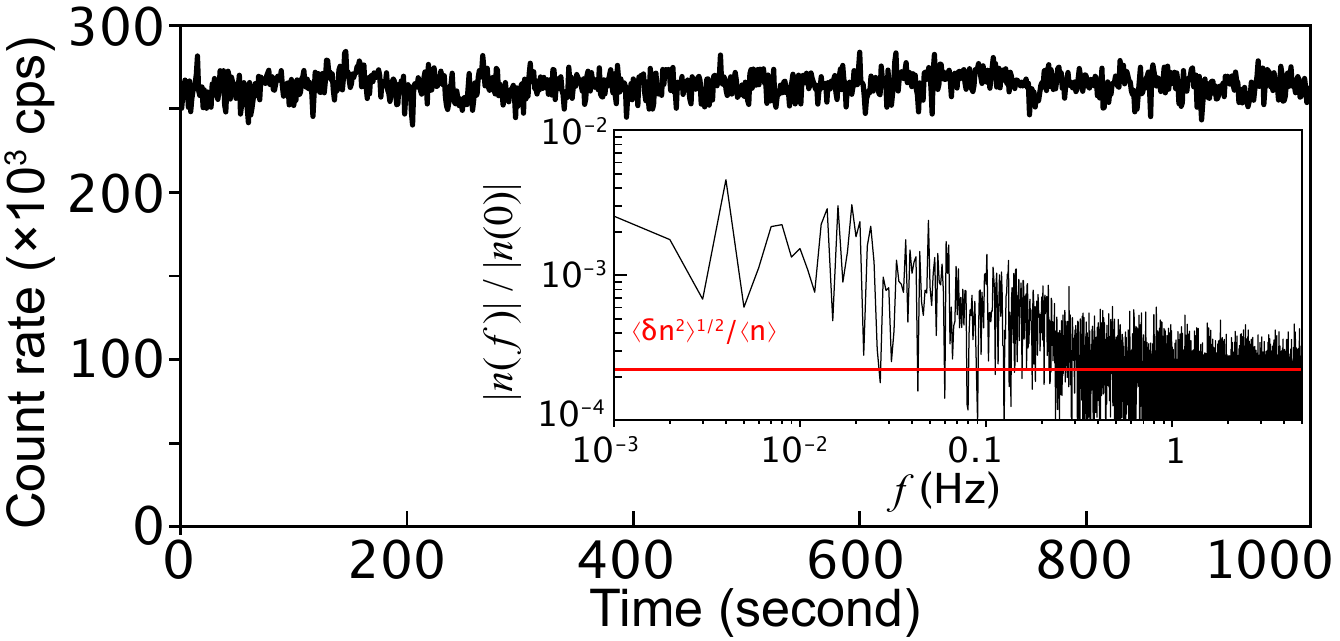}}
	\caption{ 
		Stability of the system represented by photon count rates traced over time ($n(t)$). A shot is measured for 1000 second with a sampling rate of 1 Hz.
		The original data was acquired with a sampling rate of 100 Hz and reproduced in Fourier domain ($n(f)$), and shown in the inset. The red line is a ratio of the root mean square fluctuation to the average count rate.
	}
	\label{fig:stability}
\end{figure}

Finally, we demonstrated that the stability of the system integrated from separate modules has an acceptable level compared to monolithic designs with no modularization.
We examined this by the photon count rate over time ($n(t)$) traced for a single photon stream.
An amplitude noise in the excitation beam was suppressed by a liquid crystal modulator in the \textit{excitation module}, and measurements were carried out under controlled environments of humidity and temperature.
Data shown here as in Fig. \ref{fig:stability} was acquired for 1000 seconds.
The count rate is moderate to avoid saturation of the fluorescence emission and the speed limit of SPCMs.
We kept watching photon count rate for repeated shots of measurement, and have not observed noticeable changes during $1$ hour.
A ratio of the root mean square fluctuation to the average count rate ($\langle \delta n^2\rangle^{\frac{1}{2}}/\langle n \rangle$) is 0.07 \%, and the maximum ratio of the fluctuation in Fourier domain ($|\delta n(f)|/|n(0)|$) is 0.3 \% for a bandwidth of 50 Hz, as shown in the inset of Fig. \ref{fig:stability}.
This level of stability was also helped by careful selections of mechanical components immune to vibrations. 

%%%%%%%%%%%%%%%%%%%%%%%%%%%%%%%%%%%%%%%%%%%%%%%%%%%%%%%%%%%%%%%%%%%%
%%%%%%%%%%%%%%%%%%%%%%%%%%%%%%%%%%%%%%%%%%%%%%%%%%%%%%%%%%%%%%%%%%%%
\section{Discussions}
We  implemented ideas of modular optical systems in developing the integrated system to characterize single photons and to distribute them via a single mode fiber.
Key ideas are that the optimal point of alignments is given independently to each module, and that its reference is maintained by optical apparatuses inside and in-between modules.
Those features gave great advantages to enhance the utility and to achieve a high degree of stability.
The stability of maintaining single photon count rate and the degree of fluctuations, which we demonstrated here, are in the level acceptable for precise measurements of quantum radiometry\cite{Rodiek:2017en}.
Outputs that we facilitated with single mode fibers are reserved for practical use of single photons in a broader range of applications.

The current design considers primarily high-purity and low-loss connections between modules, based on the lowest order Gaussian mode and the fundamental mode of SMF.
This, however, sacrifices collection efficiency for single photon emitters.
Non-Gaussian and higher order modes inevitably occur for dipole radiation, and they are not suitably coupled to SMF.\cite{Zschiedrich:2018fc}

We suggest two useful solution: few-mode fiber (FMF), nano-structures of optical antenna and cavity.
Additional modes in FMF allows varieties of spatial profile, and increases the collection efficiency.
The careful optimization with FMF does not harm the modal compatibility required for the modular design.
Specifically, we suggest graded-index (GI-) and photonic crystal (PC-) FMF.\cite{6683058, Sakamoto:11}
GI-FMF with collimation optics of high f-number can be employed as the spatial filter of confocal microscope, and having short transmission length without excessive bending of fiber avoids serious intermodal coupling.
PC-FMF will be promising due to the well-defined number of mode and large MFD.\cite{Sakamoto:11}
The large MFD gives the important advantage of robustness for fiber coupling as we discussed in Sec. \ref{sec:SMFcol}.

Nano-structures of optical antenna and cavity are advanced tools to reshape single photon emission.
The emission coupled with those structure can be tailored into one fitted for SMF coupling, which resulted a high extraction efficiency.\cite{PhysRevLett.122.113602,Chen:2018vy,C7NR08249E}
Those structures are imprinted on the sample surface, and expected to work without modification in our optics system.

\section*{Funding}
National Research Council of Science and Technology (NST) (CAP-15-08-KRISS); Korea Research Institute of Standards and Science (KRISS) (KRISS-2019-GP2019-0020).

\section*{Disclosures}
\medskip
\noindent The authors declare no conflicts of interest.

%\bibliography{bibliography.bib}  

\end{document}